\documentclass{article}
\usepackage[utf8]{inputenc}
\usepackage[english]{babel}
\usepackage{csquotes}
\usepackage{graphicx}
\usepackage[hidelinks]{hyperref}
\usepackage{amsmath}
\usepackage{amsfonts}
\usepackage{algorithm}
\usepackage{algpseudocode}
\usepackage{enumitem}
\usepackage{array}
\usepackage{booktabs}
\usepackage[left=2cm,right=2cm,top=2cm,bottom=2cm]{geometry}
\usepackage[
    backend=biber,
    style=numeric,
    autocite=footnote,
    sorting=nty,
    sortcase=false,
    url=true,
    date=year,
    eprint=false,
    hyperref=auto,
    maxbibnames=99,
    giveninits=true,
    isbn=false
]{biblatex}

\hypersetup{%
    pdftitle=An efficient simulation approach for the convection-dominated species transport about single rising bubbles,
    pdfauthor=Andre Weiner,
    pdfsubject=article,
    unicode=true,
    breaklinks=true,
    colorlinks=false
}

\title{Backpropagation and gradient descent for an optimized dynamic mode decomposition}
\author{Andre Weiner$^{1}$ and Richard Semaan$^{2}$ \\
        \small $^1$Technical University of Dresden, Institute of Fluid Mechanics,\\
        \small andre.weiner@tu-dresden.de, ORCID ID 0000-0001-5617-156\\
        \small $^2$Technical University of Braunschweig, Institute of Fluid Mechanics,\\
        \small r.semaan@tu-braunschweig.de, ORCID ID 0000-0002-3219-0545\\
}
\date{\today}

\providecommand{\keywords}[1]
{
  \small	
  \textbf{\textit{Keywords---}} #1
}

\begin{document}

\maketitle

\begin{abstract}
  We present a robust and flexible optimization approach for dynamic mode decomposition analysis of data with complex dynamics and low signal-to-noise ratios.
The approach borrows techniques and insights from the field of deep learning.
Specifically, we employ automatic differentiation and stochastic gradient descent to compute eigenvalues, modes, and mode amplitudes simultaneously.
The method allows embedding regularization or physical constraints into the operator definition.
The optimization approach is applied to three examples of increasing complexity, the most challenging of which is an experimental dataset of transonic shock buffet on a swept wing at realistic flight conditions.

\end{abstract}


\section{Introduction}
\label{sec:intro}
Dynamic mode decomposition (DMD) is a popular data-driven algorithm for analyzing and modeling dynamical systems.
Since its inception \cite{schmid2010}, DMD has found widespread application in the fluid dynamics community and beyond \cite{kutz2014}.
One of the main arguments in favor of DMD is its excellent interpretability.
Moreover, the algorithm is very flexible in that, at least in principle, only a sequence of observations is required to analyze and identify the system.
A limitation of the standard DMD variant is its sensitivity to nonlinearity and noise, where noise is broadly defined as unwanted effects in the data with little to no coherence.
Noise leads to a strong bias in the dynamics of DMD modes, which decay significantly faster than they should.
A mathematical justification for this trend can be found in Dawson et al. \cite{dawson2016}.
A number of DMD variants have been proposed to correct for this bias.
We refer the reader to Schmid \cite{schmid2022} for a detailed review of the different variants.
In the following, we describe our experience as practitioners with several variants and point out their upsides and shortcomings.
The mathematical details of the relevant variants are provided in the next section.

We start with the total least squares DMD (TDMD) of Hemati et al. \cite{hemati2017}, where an additional subspace projection based on an augmented data matrix is performed before the regular operator identification step.
The TDMD is straightforward to implement, cost-efficient, and mitigates the bias described above, especially when the noise level is low and the noise is Gaussian.
For more challenging datasets with resolved turbulence \cite{weiner2023}, we have found the TDMD improvement over the standard DMD to be relatively small.

Robust DMD by Scherl et al. \cite{scherl2020} is similar to the TDMD in that noise is removed from the snapshot data in a pre-processing step before the regular DMD algorithm is applied.
Noise reduction is achieved using a robust principal component analysis (RPCA) to decompose the data into low-rank and sparse contributions.
The operator identification is performed only on the low-rank contributions.
The core of the RPCA is the \textit{principal components pursuit} problem, which may be solved using a variant of the augmented Lagrange multiplier (ALM) algorithm.
RPCA also introduces a hyperparameter to balance between low-rank and sparse contributions.
The success of the ALM optimization strongly depends on the value of this sparsity control parameter.
In previous investigations, we attempted to apply robust DMD to experimental data of delta wing vortices \cite{rathje2022} and to simulation data of transonic shock buffet \cite{weiner2023}.
The algorithm converged for small values of the sparsity parameters, but the bias improvement was small.
The inexact ALM, the same algorithm recommended in Scherl et al. \cite{scherl2020}, did not converge for larger sparsity values.

The higher-order DMD (HODMD) introduced by Le Clainche et al. \cite{clainche2017} employs DMD on a data matrix augmented with time delays.
The snapshots are first projected on a truncated proper orthogonal decomposition (POD) basis to reduce the computational cost.
The HODMD significantly reduces the erroneous damping and allows the reconstruction of nonlinear dynamics.
This improvement comes at the cost of additional hyperparameters and reduced interpretability.
The main challenge lies in choosing a suitable number of time delays.
While more delays tend to improve the reconstruction accuracy, they also increase the size of DMD modes.
In principle, if the HODMD is computed with $d$ time delays, the number of potentially meaningful mode components to be analyzed also increases by $d$, because each mode has the size of $d$ snapshots.
Therefore, the problem of selecting and interpreting dynamically meaningful modes also becomes more challenging.
Moreover, the resulting dynamics tend to be sensitive to the number of delay coordinates.

One variant that significantly improves the noise bias is the optimized DMD introduced by Askham and Kutz \cite{askham2018}.
Here, the authors define the DMD operator as a much more restrictive, nonlinear optimization problem that can be solved using variable projection (VarPro).
Solving this optimization problem reduces the reconstruction error, especially for challenging datasets.
A low reconstruction error indicates that the DMD reflects the dynamics in the data well and, therefore, provides the basis for building a solid interpretation of the results.
However, focusing too much on the reconstruction error can be misleading, especially for noisy datasets.
Simple numerical experiments show that an accurate reconstruction is not necessarily the same as a precise capture of the underlying dynamics \cite{wu2021}.
The same observation is also true for HODMD.
This seemingly contradictory statement is strongly connected to the problem of overfitting and choosing a suitable effective rank, which we will discuss in more detail next.

One important hyperparameter common to all DMD variants is the size of the POD basis on which to project the DMD operator.
Truncating the POD basis is essential for several reasons.
Firstly, the truncation removes some noise in the snapshots, thereby improving the signal-to-noise ratio (SNR).
Secondly, the number of free parameters in terms of eigenvectors and eigenvalues is reduced.
A smaller number of modes and dynamics simplifies the subsequent interpretation and mitigates the chance of overfitting the operator to the data.
The latter argument is rarely discussed in DMD applications, even though overfitting is a well-known pitfall in the machine learning community.
A typical symptom of overfitting in time series forecasting is the inability to predict sequences of states beyond the data used to fit model parameters.
To emphasize this potential threat, we can take the extreme case of fitting the DMD operator to a tall-skinny matrix full of randomly sampled numbers.
When employing the full POD basis, we could obtain perfectly nonvanishing dynamics and a low reconstruction error.
Of course, predicting states beyond the last snapshot cannot yield meaningful results, and the reconstruction error would increase immediately.

Reducing the number of free parameters is only one way to avoid overfitting.
Other approaches are early stopping and further constraining the optimization problem.
Early stopping requires splitting the data into training and validation sets.
The iterative optimization is stopped once the error computed on the validation data stops decreasing or starts increasing.
Even for noniterative DMD variants, splitting the data could still help determine an appropriate size of the POD basis.
Constraining the optimization problem could be done in several ways.
Examples include the sparsity-promoting DMD of Jovanović et al. \cite{jovanovic2014} and the physics-informed DMD of Baddoo et al. \cite{baddoo2023}.
However, even with these techniques, the results are not guaranteed to be meaningful.
For example, even a small bias in the eigenvalues leads to a quickly vanishing solution.
In this scenario, evaluating the performance on a validation set is meaningless since, for time series data, one would reserve snapshots at the end of the sequence for validation.
Using the optimized DMD could provide a better validation measure, but one must also constrain the optimization to prevent overfitting.
Embedding additional constraints, such as sparsity promotion or physical constraints, into the VarPro algorithm is not trivial.

This article proposes an alternative optimization approach for fitting the DMD operator to data.
The approach borrows heavily from the deep learning (DL) community.
In particular, we obtain the DMD modes and dynamics by solving nonlinear optimization problems using stochastic gradient descent. 
Automatic differentiation (AD) yields the gradient of the loss function with respect to the free parameters (eigenvectors and eigenvalues).
The optimization is very robust and enables enormous flexibility in the definition of the DMD operator.
Note that this idea differs from the so-called deep learning DMD (DLDMD) of Alford-Lago et al. \cite{alfordlago2022}.
The latter work embeds the DMD in an encoder-decoder structure so that the potentially nonlinear dynamics of the original observations are transformed into linear ones by the encoder network.
DLDMD also requires a differentiable DMD implementation, but the DMD operator is not optimized jointly with the encoder-decoder networks.

The remainder of this article is structured as follows. 
In section \ref{sec:theory}, we briefly review the mathematical details of selected DMD variants and describe the proposed optimization in more detail.
In section \ref{sec:applications}, we present results for three different applications of increasing complexity, the most challenging of which is an experimental dataset of transonic shock buffet with a low SNR.
The complete analyses of all three test cases are also available online at \url{https://github.com/AndreWeiner/optDMD}. 
The proposed DMD variant is available as open-source software \cite{weiner2021}.

\section{Theory}
\label{sec:theory}
\subsection{Dynamic mode decomposition}
\label{sec:exact}

The starting point for DMD is a set of $N$ snapshots $\lbrace\mathbf{x}_1, \mathbf{x}_2, \ldots, \mathbf{x}_N \rbrace$.
Each snapshot $\mathbf{x}_n \in \mathbb{R}^M$ describes the state of the system at time instance $t_n = n\Delta t$.
We assume that the time increment between two consecutive snapshots, $\Delta t$, is constant, but note that this assumption could be relaxed to some extent \cite{tu2014}.
The core idea of DMD is to find a linear operator $\mathbf{A} \in \mathbb{R}^{M\times M}$ that advances the state by $\Delta t$ from $\mathbf{x}_n$ to $\mathbf{x}_{n+1}$:
\begin{equation}
\label{eq:dmd_map}
    \mathbf{x}_{n+1} = \mathbf{A} \mathbf{x}_{n}.
\end{equation}
Since it is unlikely in practice to find a single operator that exactly satisfies \eqref{eq:dmd_map} for all pairs of snapshots, $\mathbf{A}$ is defined to be optimal in the least squares sense over all pairs of snapshots:
\begin{equation}
\label{eq:least_squares}
    \underset{\mathbf{A}}{\mathrm{argmin}}\left|\left| \mathbf{Y}-\mathbf{AX} \right|\right|_F
    =\mathbf{YX}^\dagger,
\end{equation}
where $\mathbf{X} = \left[ \mathbf{x}_1, \ldots, \mathbf{x}_{N-1} \right]^T$, $\mathbf{Y} = \left[ \mathbf{x}_2, \ldots, \mathbf{x}_{N} \right]^T$, $\mathbf{X}, \mathbf{Y} \in \mathbb{R}^{M\times N-1}$, $||\ldots||_F$ denotes the Frobenius norm, and $^\dagger$ denotes the pseudo-inverse.
Computing the full operator according to \eqref{eq:least_squares} is difficult and unnecessary in practice.
Instead, a rank-$r$ approximation of $\mathbf{A}$ via a singular value decomposition (SVD) of the first snapshot matrix, $\mathbf{X}\approx \mathbf{U}_r\mathbf{\Sigma}_r\mathbf{V}_r^T$, is invoked:
\begin{equation}
    \tilde{\mathbf{A}} = \mathbf{U}_r^T\mathbf{AU}_r = \mathbf{U}_r^T \mathbf{Y} \mathbf{V}_r\mathbf{\Sigma}_r^{-1},
\end{equation}
where $\mathbf{U}\in \mathbb{R}^{M\times r}$, $\mathbf{V} \in \mathbb{R}^{N-1\times r}$, $\mathbf{\Sigma} \in \mathbb{R}^{r\times r}$, and $\tilde{\mathbf{A}} \in \mathbb{R}^{r\times r}$.
Assuming that $\mathbf{A}$ and $\tilde{\mathbf{A}}$ are similar, the eigendecomposition of $\tilde{\mathbf{A}}=\mathbf{W}\mathbf{\Lambda}_r\mathbf{W}^{-1}$ yields the same nonzero eigenvalues as $\mathbf{A}=\mathbf{\Phi\Lambda\Phi}^{-1}$.
The first $r$ eigenvectors of $\mathbf{A}$ can be reconstructed according to Tu et al. \cite{tu2014}:
\begin{equation}
    \mathbf{\Phi}_r = \mathbf{Y} \mathbf{V}_r \mathbf{\Sigma}_r^{-1}\mathbf{W},
\end{equation}
where $\mathbf{\Phi}_r \in \mathbb{C}^{M\times r}$.
Alternatively, one could also reconstruct the modes according to the original algorithm by Schmid \cite{schmid2010}, yielding the so-called projected DMD modes, $\bar{\mathbf{\Phi}}_r = \mathbf{U}_r\mathbf{W}$.
In practice, there is very little difference between the exact and projected modes.
Given an approximate eigendecomposition $\mathbf{A}\approx \mathbf{\Phi}_r\mathbf{\Lambda}_r \mathbf{\Phi}_r^{\dagger}$, the state at any time instance $t_n$ with $n \in \mathbb{Z}^+$ can be easily reconstructed as:
\begin{equation}
    \label{eq:rec_n}
    \mathbf{x}_n \approx \mathbf{\Phi}_r\mathbf{\Lambda}_r^{n-1}\mathbf{\Phi}_r^{\dagger}\mathbf{x}_1.
\end{equation}
The last product in \eqref{eq:rec_n} is typically abbreviated as $\mathbf{b}_r = \mathbf{\Phi}_r^{\dagger}\mathbf{x}_1$.
The entries in $\mathbf{b}_r$ are often referred to as mode amplitudes and provide a measure of the importance of individual modes.
By introducing the diagonal matrix $\mathbf{D}_\mathbf{b} \in \mathbb{C}^{r\times r}$, with the diagonal elements given by $\mathbf{b}_r$, and the Vandermode matrix $\mathbf{V}_{\mathbf{\lambda}} = \left[\mathbf{\lambda}_r^0, \mathbf{\lambda}_r^1, \ldots, \mathbf{\lambda}_r^{N-1}\right]^T$, where $\mathbf{\lambda}_r \in \mathbb{C}^r$ are the diagonal elements of $\mathbf{\Lambda}_r$, we can approximate the full dataset $\mathbf{M}=\left[\mathbf{x}_1, \mathbf{x}_2, \ldots, \mathbf{x}_N\right]^T$ as:
\begin{equation}
    \label{eq:rec_full}
    \mathbf{M}\approx \mathbf{\Phi}_r\mathbf{D}_\mathbf{b}\mathbf{V}_\mathbf{\lambda}.
\end{equation}
Based on \eqref{eq:rec_full}, we define the normalized reconstruction error:
\begin{equation}
    \label{eq:e_rec}
    E_\mathrm{rec} = \frac{||\mathbf{M} - \mathbf{\Phi}_r\mathbf{D}_\mathbf{b}\mathbf{V}_\mathbf{\lambda}||_F}{||\mathbf{M}||_F}.
\end{equation}
For completeness, we note that the frequency $f$ and the growth rate $\xi$ of mode $i$ are computed as \cite{kutz2014}:
\begin{equation}
    \label{eq:freq}
    f_i = \frac{\Im(\mathrm{log}(\lambda_i))}{2\pi\Delta t}\quad\text{and}\quad
    \xi_i = \frac{\Re(\mathrm{log}(\lambda_i))}{\Delta t},
\end{equation}
where $\Re (\dots)$ and $\Im (\dots)$ are the real and imaginary parts, respectively.

\subsection{DMD extensions}
\label{sec:dmd_ext}

Computing the amplitudes $\mathbf{b}_r$ based on the first snapshot leads to a reconstruction that tends to be more accurate for the initial snapshots.
Many DMD variants and extensions have been introduced to alleviate this problem.
Jovanovi\'c et al. \cite{jovanovic2014} introduced an alternative linear-algebraic approach to determine $\mathbf{b}_r$, which solves an additional least squares problem:
\begin{equation}
\label{eq:opt_amp}
    \underset{\mathbf{b}_r}{\mathrm{argmin}} \left|\left|
\mathbf{M}-\mathbf{\Phi}_r\mathbf{D}_\mathbf{b}\mathbf{V}_{\mathbf{\lambda}} \right|\right|_F,
\end{equation}
The optimized amplitudes yield a more balanced error distribution over time and a lower overall reconstruction error according to \eqref{eq:e_rec}.
We emphasize that this extension only affects the reconstruction error and the mode selection, but not the linear operator or its eigendecomposition.
The same authors also provide a variant of problem \eqref{eq:opt_amp} with additional regularization:
\begin{equation}
\label{eq:spdmd}
    \underset{\mathbf{b}_r}{\mathrm{argmin}}\left(\left|\left| \mathbf{M}-\mathbf{\Phi}_r\mathbf{D}_\mathbf{b}\mathbf{V}_{\mathbf{\lambda}} \right|\right|_F+ \gamma_0\left|\left|\mathbf{b}_r\right|\right|_1\right),
\end{equation}
where $\gamma_0$ is a hyperparameter that controls the sparsity of $\mathbf{b}_r$.
The iterative solution of problem \eqref{eq:spdmd} leads to the so-called sparsity-promoting DMD (spDMD).
Sparsity promotion can be tremendously helpful in selecting dynamically important modes when the true rank of $\mathbf{A}$ is not known, which is usually the case.
Moreover, regularization is an important measure to prevent overfitting.

Another notable DMD variant was introduced by Askham and Kutz \cite{askham2018}, who proposed an optimized definition of the full DMD operator:
\begin{equation}
\label{eq:optDMD}
    \underset{\mathbf{\lambda}_r,\mathbf{\Phi}_\mathbf{b}}{\mathrm{argmin}}\left|\left| \mathbf{M}-\mathbf{\Phi}_\mathbf{b}\mathbf{V}_{\mathbf{\lambda}} \right|\right|_F,
\end{equation}
where $\mathbf{\Phi}_\mathbf{b} = \mathbf{\Phi}_r\mathbf{D}_\mathbf{b}$ are the amplitude-scaled DMD modes.
The modes and mode amplitudes are recovered by rescaling the optimized DMD modes to unit length, i.e., $b_{i} = ||\mathbf{\varphi}_{\mathbf{b},i}||_2$ and $\mathbf{\varphi}_i = \mathbf{\varphi}_{\mathbf{b},i} / b_i$.
Note that problem \eqref{eq:optDMD} is much more restrictive and difficult to solve than the original operator definition \eqref{eq:least_squares}.
While the standard operator minimizes the error over pairs of snapshots, the optimized DMD problem \eqref{eq:optDMD} considers the error propagation over time via the Vandermode matrix.
Of course, the resulting optimization problem is nonlinear and nonconvex, and must be solved by a combination of variable projection and the Levenberg/Levenberg-Marquardt algorithm, which is not guaranteed to converge to the global optimum.
However, even local optima of \eqref{eq:optDMD} significantly reduce the negative impact of noise \cite{askham2018}.

Extensions \eqref{eq:opt_amp} and \eqref{eq:optDMD} improve reconstruction error and mode selection.
The latter can be further stabilized by considering the integral contribution of each mode eigenvalue pair over time, as suggested by Kou and Zang \cite{kou2017}:
\begin{equation}
\label{eq:int_criterion}
    I_i = \sum\limits_{j=0}^{N-1} |b_i\lambda_i^j|\,,
\end{equation}
where $b_i$ and $\lambda_i$ denote the $i$th component of $\mathbf{b}_r$ and $\mathbf{\lambda}_r$, respectively.
Note that, compared to \cite{kou2017}, we omitted the multiplication by $\Delta t$ and $||\mathbf{\varphi}_i||_2$ since both values do not affect the mode sorting.

\subsection{Backpropagation and gradient descent for an optimized DMD}
\label{sec:bpdmd}

Over the past decade, the DL community has developed sophisticated optimization techniques for the training of deep neural networks.
The free parameters of such networks are almost exclusively optimized using a combination of stochastic gradient descent and automatic differentiation (AD).
AD greatly simplifies the computation of the partial derivatives of any complex function with respect to its inputs. 
The only strict requirement is that the function can be decomposed into a sequence of simple functions with known derivatives.
Applying the chain rule to this sequence of base functions yields the partial derivatives of the complex function with respect to the free parameters.
Modern DL libraries such as PyTorch or TensorFlow make AD accessible.
In fact, many data-driven algorithms that do not necessarily require AD can also benefit from a differentiable implementation.
An example is GPyTorch \cite{gardner2018}, an implementation of Gaussian processes in PyTorch.
GPyTorch uses AD and gradient descent to optimize the hyperparameters of Gaussian processes.
This gradient-based optimization is much more efficient than random or grid search, and is possible because all operations involved in a Gaussian process computation can be expressed as basic PyTorch operations.
The same concept is employed in the DLDMD \cite{alfordlago2022} variant mentioned in the introduction.
Given a differentiable DMD implementation, the DMD reconstruction can be used in the loss function of an autoencoder.
By optimizing the weights of both the encoder and decoder networks, the authors obtained transformed observables whose dynamics are linear.

In this paper, we generalize the DMD operator definition to:
\begin{equation}
\label{eq:optDMDgen}
    \underset{\mathbf{\lambda}_r,\mathbf{\Phi}_\mathbf{b}}{\mathrm{argmin}}
    \underbrace{
    \left(\left|\left| \mathbf{M}-\mathbf{\Phi}_b\mathbf{V}_{\mathbf{\lambda}} \right|\right|_F +
    f(\mathbf{\lambda}_r,\mathbf{\Phi}_\mathbf{b})\right)}_{L},
\end{equation}
and propose solving it using stochastic gradient descent combined with AD.
The function $f(\mathbf{\lambda}_r,\mathbf{\Phi}_\mathbf{b})$ is a placeholder to include soft constraints.
Some examples of $f(\mathbf{\lambda}_r,\mathbf{\Phi}_\mathbf{b})$ include:
\begin{align}
    \label{eq:f_opt} &\text{optimized DMD} &f(\mathbf{\lambda}_r,\mathbf{\Phi}_\mathbf{b}) &= 0\\
    &\text{Tikhonov regularization} &f(\mathbf{\lambda}_r,\mathbf{\Phi}_\mathbf{b}) &= \label{eq:f_reg} \gamma_0 ||\mathbf{b}_r||_2\quad \text{with}\quad b_{i} = ||\mathbf{\varphi}_{\mathbf{b},i}||_2 \\
    \label{eq:f_orth} &\text{orthogonal operator} & f(\mathbf{\lambda}_r,\mathbf{\Phi}_\mathbf{b}) &=
    \gamma_0 ||\mathbf{AA}^T - \mathbf{I}||_F\quad \text{with}\quad \mathbf{A} =
    \mathbf{\Phi}_r\mathbf{\Lambda}_r \mathbf{\Phi}_r^{\dagger}\\
    \label{eq:f_symm} &\text{symmetric operator} & f(\mathbf{\lambda}_r,\mathbf{\Phi}_\mathbf{b}) &=
    \gamma_0 ||\mathbf{A} - \mathbf{A}^T||_F\quad \text{with}\quad \mathbf{A} =
    \mathbf{\Phi}_r\mathbf{\Lambda}_r \mathbf{\Phi}_r^{T}
\end{align}
In \eqref{eq:f_orth}, $\mathbf{I} \in \mathbb{R}^{M\times M}$ is the identity matrix.
As described by Baddoo et al. \cite{baddoo2023}, the orthogonality and symmetry constraints may be interpreted in a physical context, given a suitable state vector.
For example, if the norm of the state yields a conserved quantity, the orthogonality constraint enforces energy conservation (in an integral sense).
To make the solution of \eqref{eq:f_orth} or \eqref{eq:f_symm} computationally tractable, one can avoid reconstructing the full $M\times M $ operator by projecting the data matrix $\mathbf{M}$ onto a reduced basis before performing the DMD.
With the examples \eqref{eq:f_opt}-\eqref{eq:f_symm}, we aim to emphasize the flexibility of the proposed algorithm.
Our implementation allows passing $f(\mathbf{\lambda}_r,\mathbf{\Phi}_\mathbf{b})$ as an argument to the optimization routine.
Thanks to AD, no other modifications are needed in the rest of the algorithm. In section \ref{sec:applications}, we restrict ourselves to the variants \eqref{eq:f_opt} and \eqref{eq:f_reg}.

To solve problem \eqref{eq:optDMDgen}, we employ a generic DL optimization routine with learning rate scheduling and early stopping similar to the one described in chapter 5 of reference \cite{raff2022}.
The main steps are outlined in algorithm \ref{alg:optDMD}.
We use the eigenvectors and eigenvalues resulting from the exact DMD to initialize the optimization.
Optimizing all eigenvalue-eigenvector pairs could violate the complex conjugate pairs constraint for real input data.
Therefore, only eigenvalues with a positive imaginary part and the corresponding modes are retained and optimized.
The conjugate complex eigenvalues or eigenvectors can be computed straightforwardly if needed.
We split the snapshots into a training set and a validation set.
By default, the first $75\%$ of snapshots are used for training, while the remaining snapshots are used for testing.
Optionally, the training set can be further divided into randomly drawn batches of size $N_\text{batch}$.
The batches are drawn without replacement so that all snapshots are processed once every epoch $e$.
Batch gradient descent is a common practice in DL and can significantly speed up optimization.
While the computational effort per epoch increases, the total number of epochs required is typically smaller due to the additional updates in each epoch.

Small modifications to the loss function \eqref{eq:optDMDgen} are required when training on batches $D_\text{batch}$.
First, the reconstruction error is computed only for snapshots $\mathbf{x}_n \in D_\text{batch}$.
Second, we normalize the Frobenius norm with the total number of elements contained in a batch, i.e., $MN_\text{batch}$.
Employing the normalized norm reduces the need to adjust the learning rate between different datasets.
It also enables comparison between loss values computed with and without batch training.
To update the parameters $\theta= \lbrace \mathbf{\Phi}_\mathbf{b}, \mathbf{\lambda}_r \rbrace$, we use the ADAM optimizer \cite{kingma2017}.
ADAM is a first-order gradient descent algorithm with several empirical extensions.
While ADAM adjusts the update rate for each parameter in $\theta$ based on the history of the gradient $g = \partial L /\partial \theta$, it is also beneficial to reduce the learning rate $\alpha$ during the optimization.
Here we employ a plateau-based reduction of the learning rate \cite{raff2022}.
Once the validation loss does not decrease within a given number of epochs, the learning rate is halved.
The validation loss is evaluated similarly to the convergence criterion.
The optimization is terminated when the validation loss does not decrease for a given number of epochs.
For the special case that $D_{\text{val}} = \emptyset$, the training loss is evaluated in both the learning rate adjustment and the convergence check.
Note that the full implementation details are publicly available in the online repository \url{https://github.com/AndreWeiner/optDMD}.

\begin{algorithm}
\caption{Optimization of eigenvectors and eigenvalues with batch gradient descent, learning rate adjustment, and early stopping.}
\label{alg:optDMD}
\begin{algorithmic}
\State \textbf{Input:} dataset $D$, initial learning rate $\alpha_0$, batch size $N_{\text{batch}}$, iteration maximum $e_{\text{max}}$, training data size $N_{\text{train}}$, loss function $L$
\State \textbf{Output:} optimal eigenvectors and values $\mathbf{\Phi}_\mathbf{b}, \mathbf{\lambda}_r$
\State Initialize parameters $\theta = \lbrace \mathbf{\Phi}_\mathbf{b}, \mathbf{\lambda}_r \rbrace$ with exact DMD
\State Initialize learning rate $\alpha \gets \alpha_0$
\State Initialize iteration count $e \gets 0$
\State Initialize stopping flag $stop \gets False$
\State Create training/validation split $D_{\text{train}} = \lbrace \mathbf{x}_n \in D\ |\ n \leq N_{\text{train}} \rbrace$, $D_{\text{val}} = \lbrace \mathbf{x}_n \in D\ |\ n > N_{\text{train}} \rbrace$
\While{$e < e_{\text{max}}$ and not $stop$}
\State $L_{\text{train}} \gets 0$
\For{$i \gets 1$ to $\lceil N_\text{train}/N_\text{batch}\rceil$}
  \State $D_{\text{batch}} \gets \mathrm{RandomSample}(D_{\text{train}}, N_\text{batch})$ \Comment{draw without replacement}
  \State $L_{\text{batch}} \gets L(D_\text{batch})$ \Comment{batch loss}
  \State $g_\text{batch} \gets \mathrm{Backpropagation}(L_{\text{batch}})$ \Comment{batch gradient}
  \State $\theta \gets \mathrm{ADAM}(\theta, \alpha, g_\text{batch})$ \Comment{gradient descent}
  \State $L_\text{train}\gets L_\text{train} + L_{\text{batch}}$
\EndFor
    \State $L_{\text{val}} \gets L(D_\text{val})$ \Comment{validation loss}
    \State $\alpha \gets \mathrm{ReduceLearningRate}(\alpha, L_{\text{train}}, L_{\text{val}})$ 
    \State $stop \gets \mathrm{EarlyStopping}(L_{\text{train}}, L_{\text{val}})$ \Comment{check for stall, overfitting}
    \State $e \gets e + 1$
\EndWhile
\State \textbf{Return} optimal $\mathbf{\Phi}_\mathbf{b}, \mathbf{\lambda}_r$
\end{algorithmic}
\end{algorithm}

\section{Results}
\label{sec:applications}
In this section, we compare the optimized DMD obtained via the procedure introduced in section \ref{sec:bpdmd} (abbreviated as ADAM) to the exact DMD and the optimized DMD obtained via VarPro.
Moreover, we also present results for the ADAM variant with an additional POD subspace projection step (abbreviated as ADAM/POD).
The implementation characteristics are summarized in table \ref{tab:dmd_variants}.
The variant with subspace projection is beneficial for large and noisy state vectors.
In a preprocessing step, the snapshots are projected onto the first $r$ left-singular vectors of $\mathbf{M}$.
Given the decomposition $\mathbf{M}=\hat{\mathbf{U}}\hat{\mathbf{\Sigma}}\hat{\mathbf{V}}^T$, the projected data matrix becomes $\hat{\mathbf{M}} = \hat{\mathbf{U}}_r^T\mathbf{M}$, with $\hat{\mathbf{M}} \in \mathbb{R}^{r\times N}$.
The operator fitting is then performed on $\hat{\mathbf{M}}$.
The full-state modes can be recovered as $\mathbf{\Phi}_r \approx \hat{\mathbf{U}}_r \hat{\mathbf{\Phi}}_r$.
Similarly, the reconstruction of the full dataset is achieved by $\mathbf{M} \approx \hat{\mathbf{U}}_r \hat{\mathbf{\Phi}}_r\hat{\mathbf{D}}_\mathbf{b}\hat{\mathbf{V}}_\mathbf{\lambda}$.
Note that performing the optimized DMD on the POD coefficients is equivalent to solving problem \eqref{eq:optDMD} for the rank-$r$ approximation $\mathbf{M}_r = \hat{\mathbf{U}}_r\hat{\mathbf{\Sigma}}_r\hat{\mathbf{V}}_r^T$.
For more details, we refer the reader to algorithm 3 and proposition 1 in Askham et al. \cite{askham2018}.
The same technique is also employed in the HODMD \cite{clainche2017} to alleviate the additional cost due to the time-delay embedding.

In the following, we assess the proposed approach on three problems of increasing difficulty: a one-dimensional analytical signal, a numerical simulation of a flow over a circular cylinder, and experimental surface pressure measurements over a wing undergoing a transonic buffet.

\begin{table}[htbp]
    \centering
    \begin{tabular}{m{3cm} m{8cm}}
         abbreviation & implementation characteristics \\\toprule
         exact DMD & \begin{itemize}[noitemsep,topsep=0pt]
             \item DMD according to section \ref{sec:exact}
             \item optimization of $\mathbf{b}_r$ according to \eqref{eq:opt_amp}
             \item flowTorch implementation \cite{weiner2021}\vspace{-\baselineskip}
         \end{itemize} \\\midrule
         ADAM & \begin{itemize}[noitemsep,topsep=0pt]
             \item optimized DMD according to section \ref{sec:bpdmd}
             \item flowTorch implementation \cite{weiner2021}\vspace{-\baselineskip}
         \end{itemize} \\\midrule
         ADAM/POD & \begin{itemize}[noitemsep,topsep=0pt]
             \item initial projection $\hat{\mathbf{M}} = \hat{\mathbf{U}}_r^T\mathbf{M}$
             \item same optimization as ADAM
             \item reconstruction $\mathbf{M} \approx \hat{\mathbf{U}}_r \hat{\mathbf{\Phi}}_r\hat{\mathbf{D}}_\mathbf{b}\hat{\mathbf{V}}_\mathbf{\lambda}$\vspace{-\baselineskip}
         \end{itemize} \\\midrule
         VarPro & \begin{itemize}[noitemsep,topsep=0pt]
             \item optimized DMD according to reference \cite{askham2018}
             \item PyDMD implementation \cite{demo2018}\vspace{-\baselineskip}
         \end{itemize}\\\bottomrule
    \end{tabular}
    \caption{Implementation characteristics of the tested DMD variants.}
    \label{tab:dmd_variants}
\end{table}

\subsection{One-dimensional analytical example}
\label{sec:1d_example}

The first problem is example 2 in Askham and Kutz \cite{askham2018}.
The signal consists of two translating sinusoids:
\begin{equation}
    \label{eq:sin_signal}
    z(x,t) = \mathrm{sin}(k_1x-\omega_1t)e^{\gamma_1t} +  \mathrm{sin}(k_2x-\omega_2t)e^{\gamma_2t} + \sigma s,
\end{equation}
where $k_1=1$, $\omega_1 = 1$, $\gamma_1 = 1$, $k_2 = 0.4$, $\omega_2=3.7$, and $\gamma_2=-0.2$.
The first component grows rapidly, while the second one decays slowly.
The last term corrupts the data with noise sampled from a normal distribution with zero mean and unit standard deviation, $s\sim \mathcal{N}(0, 1)$.
The noise is scaled by a constant factor $\sigma$ to allow the study of varying SNR.
Function \eqref{eq:sin_signal} is evaluated at 300 equidistant points $0 \le x \le 15$ and 512 equidistant times $0 \le t \le 2\pi$ to generate the dataset.
In this domain, the value range of signal one is about two orders of magnitude larger than that of signal two.
Therefore, recovering the two exact eigenvalue pairs of the signal $\gamma_1\pm i\omega_1$ and $\gamma_2\pm i\omega_2$ becomes increasingly difficult with decreasing SNR.

\begin{figure}[htbp]
    \centering
    \includegraphics[width=0.8\textwidth]{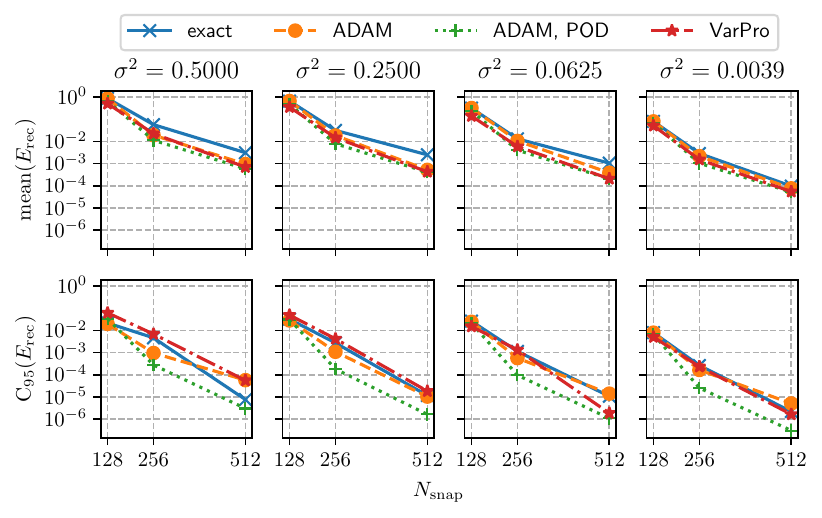}
    \caption{One-dimensional analytical example: mean and $95\%$ confidence interval of the reconstruction error \eqref{eq:e_rec} for different noise levels $\sigma$ and numbers of snapshots $N_\mathrm{snap}$.}
    \label{fig:rec_err_simple}
\end{figure}

For all DMD variants, the truncation rank is set to $r=4$.
Since the true rank of the data is known, there is no need to split the data into training and validation sets.
All of the data are used for training.
As it is common practice in machine learning, we normalize the data to the range $\left[-1, 1\right]$.
No additional constraints are used in the ADAM variants, i.e., $f(\mathbf{\Phi}_r, \mathbf{\lambda}_r)=0$, yielding the solution to problem \eqref{eq:optDMD}.
The exact DMD yields a very accurate initialization for this simple example.
Therefore, we reduce the initial learning rates to $\alpha_0 \in \lbrace 5\times 10^{-5}, 1\times 10^{-5}\rbrace$ for the ADAM variants (lower learning rates for lower $\sigma$).
Otherwise, the optimization will take unnecessarily long for low noise levels.
The VarPro variant is also initialized with the eigenvalues resulting from the exact DMD.
Each numerical test is repeated 100 times with different noise samples $s$, allowing to report means and confidence intervals of the error metrics.

Figure \ref{fig:rec_err_simple} shows the reconstruction error \eqref{eq:e_rec} for varying noise levels $\sigma^2 \in \lbrace 2^{-1}, 2^{-2}, 2^{-4}, 2^{-8}\rbrace$ and dataset sizes of $N_\mathrm{snap} \in \lbrace 128, 256, 512\rbrace$ snapshots.
The error is evaluated on the entire dataset, even if the operator identification is performed on fewer snapshots.
We compute the error against the clean dataset.
The achieved mean VarPro reconstruction error is in close agreement with the results presented in \cite{askham2018}.
All variants solving the nonlinear optimization problem \eqref{eq:optDMD} yield comparable reconstruction errors.
The exact DMD performs slightly worse for the three highest noise levels.
Moreover, the difference between the exact DMD and the other variants increases with the number of snapshots.
The $95\%$ confidence interval is about one order of magnitude smaller than the corresponding error values, indicating that the trends are statistically significant.
The ADAM/POD variant gives the smallest variance.
There are several reasons for this behavior.
First, projecting the noisy data onto the POD modes significantly reduces the noise level.
Thus, the optimization is performed on a cleaner version of the data.
Second, optimization problems typically become harder to solve as the degree of freedom increases.
The number of weights is $M\times r + r = 400\times 4 + 4 = 1204$ for ADAM and only $r\times r +r = 4\times 4 + 4 = 20$ for ADAM/POD.
Hence, the ADAM/POD optimization is more likely to be successful.

\begin{figure}[htbp]
    \centering
    \includegraphics[width=0.8\textwidth]{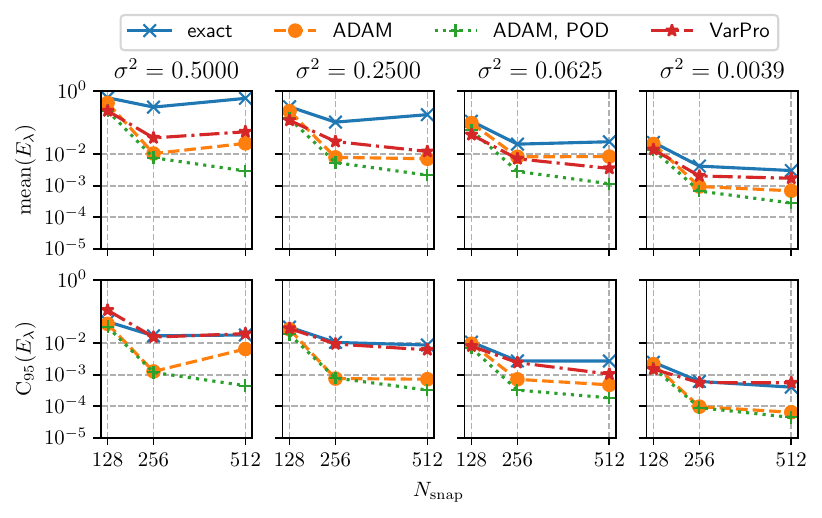}
    \caption{One-dimensional analytical example: mean and $95\%$ confidence interval of the eigenvalue prediction error \eqref{eq:ev_err} for different noise levels $\sigma$ and number of snapshots $N_\mathrm{snap}$.}
    \label{fig:ev_err_simple}
\end{figure}

For this example, the true eigenvalues $\lambda_{\mathrm{true}}$ are known.
Therefore, we can also quantify the error of the eigenvalues/dynamics, which we define as:
\begin{equation}
    \label{eq:ev_err}
    E_\mathbf{\lambda} = ||\mathbf{\lambda}_\mathrm{true} - \lambda_r||_2\,,
\end{equation}
and present in figure \ref{fig:ev_err_simple}.
Note that in the implementation, we sort the eigenvalues before computing the error, since the natural order of the computed eigenvalues may vary between variants and trials.
Interestingly, the eigenvalue error shows a different picture than the reconstruction error.
Increasing the dataset from 256 to 512 snapshots yields vanishing returns. 
For the worst SNR, the error even increases for all variants except ADAM/POD.
The main factors for this behavior are the decreasing SNR with increasing $\sigma$ and the dominance of the growing mode with increasing $N_\mathrm{snap}$ (we always take the first $N_\mathrm{snap}$ snapshots of the full dataset to perform the DMD).
The exact DMD performs significantly worse than the optimized DMD variants for all noise levels.
The ADAM/POD variant performs the best in all scenarios.
Again, the main reason is the improved SNR due to the initial projection step.
Comparing the confidence intervals of VarPro and the ADAM variants, the VarPro results appear to be much more sensitive to noise.
Of course, the good performance of the ADAM variants comes at a higher computational cost.
ADAM takes between $0.7s$ and $5.6s$ to converge, depending on the number of snapshots, while ADAM/POD takes between $0.9s$ and $9.5s$.
Note that the timing includes the SVD step required to compute the projected data and the reconstruction.
For larger snapshots, the POD variant becomes much more efficient.
In this example, the SVD step dominates the computational cost.
In addition, ADAM/POD requires more epochs to converge and achieves a lower final loss than ADAM.
For comparison, the exact DMD requires between $0.005s$ and $0.02$, while VarPro requires $0.09s$ to $0.3s$.
Note that these numbers are not the results of rigorous benchmarks but should serve as an indicator of the computational effort.

\subsection{Flow past a circular cylinder}
\label{sec:cylinder}

The second example is the flow past a circular cylinder at a Reynolds number of $Re = U_\mathrm{in}d/\nu=100$, where $U_\mathrm{in}$ is the mean inlet velocity, $d$ the cylinder diameter, and $\nu$ the kinematic viscosity.
The simulation data are part of the flowTorch dataset collection \cite{weiner2021}.
A total of 241 snapshots of the out-of-plane vorticity component $\omega_z$ are available.
Each snapshot contains the vorticity at 7190 points around the cylinder.
The flow is characterized by periodic vortex shedding with a dimensionless frequency of $S_r = fd/U_\mathrm{in} = 0.3$.
The snapshots are available at intervals of $\Delta \tilde{t}=\Delta t U_\mathrm{in}/d=0.25$ and cover about 18 vortex shedding cycles.

\begin{figure}[htbp]
    \centering
    \includegraphics[width=0.8\textwidth]{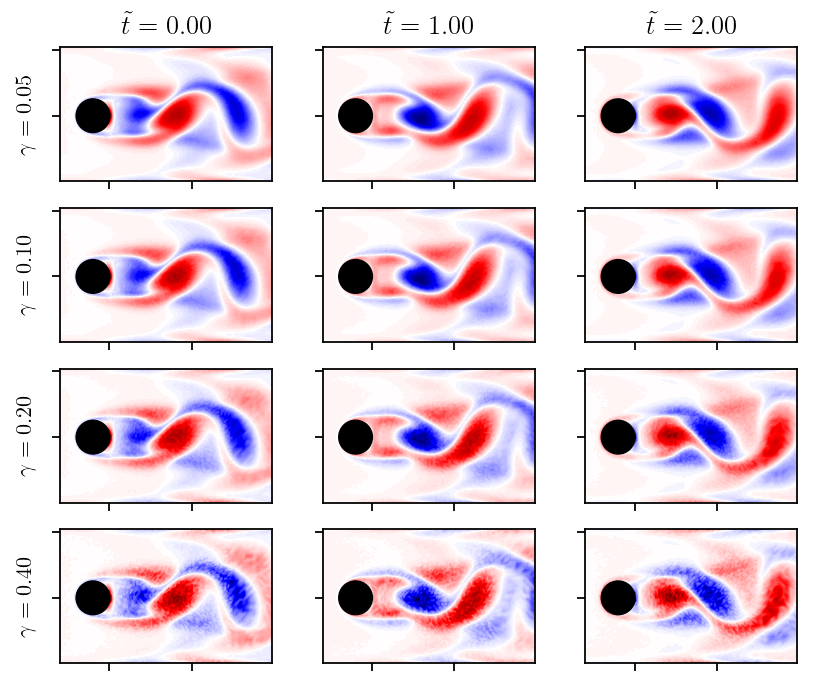}
    \caption{Sequence of three corrupted snapshots at time intervals of $4\Delta t$ with different noise levels $\gamma$. The dimensionless time is $\tilde{t}=tU_\mathrm{in}/d$. The vorticity is scaled to the range $[-1, 1]$ with red and blue colors indicating positive and negative vorticity, respectively.}
    \label{fig:cylinder_noise}
\end{figure}

For this example, we increase the complexity of the data corruption in two ways. 
First, much lower SNRs are tested, and second, the noise is not uniform but scaled by the vorticity.
Specifically, a corrupted state vector $\tilde{\mathbf{x}}_n$ is generated according to:
\begin{equation}
    \label{eq:noise_cylinder}
    \tilde{\mathbf{x}}_n = \mathbf{x}_n + \gamma |\mathbf{x}_n| \mathbf{s}_n,
\end{equation}
where $|\mathbf{x}_n|$ denotes a vorticity vector with the elementwise absolute value of $\mathbf{x}_n$, and $\mathbf{s}_n$ is a vector of random numbers sampled from a uniform distribution, i.e., $s_{n,i}\sim U(-1, 1)$.
The scalar $\gamma$ controls the SNR.
Values of $\gamma \in \lbrace 0.05, 0.1, 0.2, 0.4 \rbrace$ are tested.
Figure \ref{fig:cylinder_noise} shows a sequence of three snapshots with varying noise levels.

This example aims to test the sensitivity of different DMD variants to rank truncation and SNR.
As before, the data are not split into training and validation sets for the numerical experiments presented.
The truncation rank determines the number of adjustable parameters (number of modes and eigenvalues).
The more free parameters, the higher the chances of approximating the noise rather than the vortex shedding.
However, in a preliminary study, we ensured that the investigated truncation ranks of $r = \lbrace 10, 15, 20, 25 \rbrace$ do not lead to overfitting by comparing the reconstruction errors computed on training and validation sets ($75/25\%$ split).
We do not include these results here for brevity, but they are available in the online repository.
The initial learning rate for the ADAM variants is set to $\alpha_0 = 10^{-4}$.
The VarPro eigenvalues are initialized with those from the exact DMD computation.

\begin{figure}[htbp]
    \centering
    \includegraphics[width=0.8\textwidth]{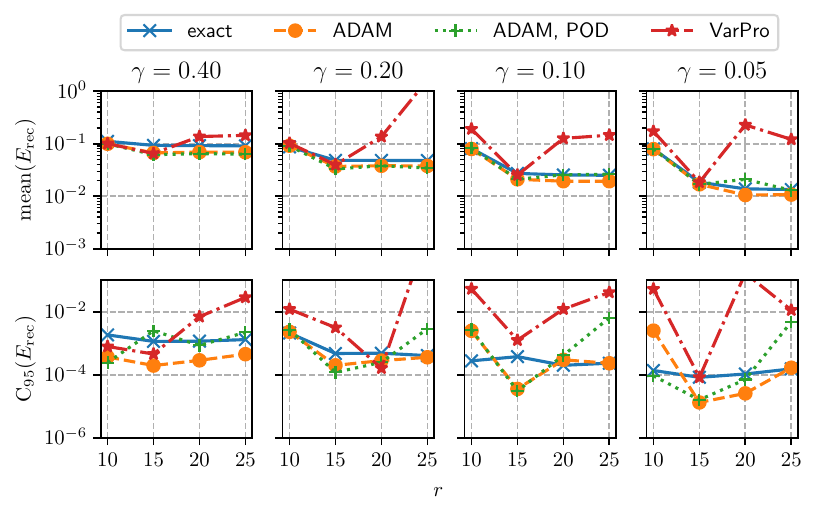}
    \caption{Cylinder flow example: mean reconstruction error (equation \eqref{eq:e_rec}) and $95\%$ confidence interval for varying noise levels $\gamma$ and truncation ranks $r$.}
    \label{fig:rec_err_cylinder}
\end{figure}

Figure \ref{fig:rec_err_cylinder} shows the mean reconstruction error over 10 trials. The error is computed against the clean data.
The confidence interval is at least one order of magnitude smaller than the reconstruction error, indicating that 10 trials are sufficient.
Note that the VarPro variant has difficulty converging, especially when the rank parameter is increased beyond $r=15$.
There is also a sensitivity to the parity of $r$, i.e., odd ranks work better than even ranks for the cylinder flow.
We varied the VarPro settings of the referenced implementation, such as initialization, constraints, and other optimization settings.
The results presented here are the best we could obtain.
VarPro experts may be able to modify the algorithm and achieve a better performance.
Nevertheless, there is an apparent lack of robustness to the rank parameter.

For the other variants, the reconstruction error increases with increasing noise level.
Furthermore, there is a vanishing return when increasing the rank parameter beyond $r=15$.
Only for the best SNR, $\gamma=0.05$, the error continues to decrease.
In terms of accuracy, all three variants perform similarly, with a slight advantage of the ADAM variants over the exact DMD.
For the largest rank, the ADAM variants show an increased confidence interval.
We suspect this effect is due to the increased number of free parameters and could be improved by lowering the learning rate.

\begin{figure}[htbp]
    \centering
    \includegraphics[width=0.8\textwidth]{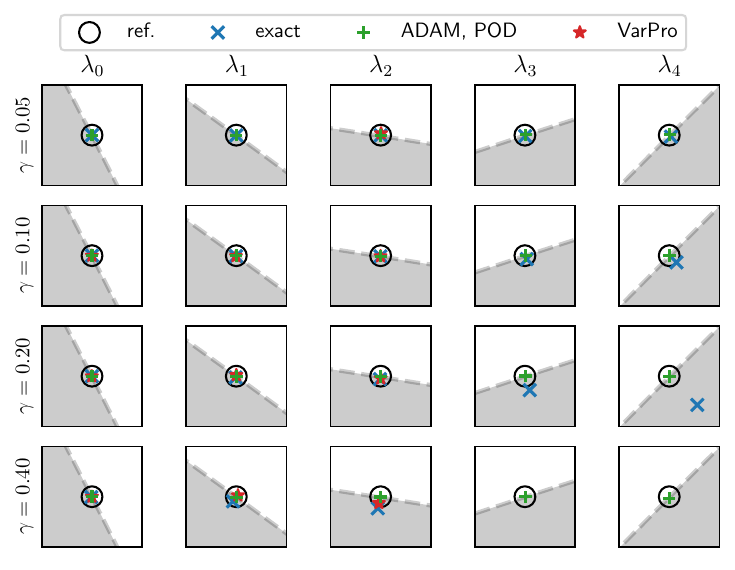}
    \caption{Cylinder flow example: comparison of the five most dominant eigenvalues (sorted by amplitude) for different noise levels $\gamma$. The truncation rank is $r=20$. The exact DMD applied to the clean data is used as a reference. For clarity, we have omitted the ADAM markers because the results were nearly identical to ADAM/POD. Each subfigure has a width and height of $0.01$.}
    \label{fig:cylinder_eigvals}
\end{figure}

As the previous example shows, low reconstruction errors are not a sufficient criterion for accurately capturing the underlying dynamics.
Figure \ref{fig:cylinder_eigvals} presents the sensitivity of the five most dominant eigenvalues to noise.
The eigenvalues are ordered by their amplitudes $\mathbf{b}_r$.
The eigenvalue $\lambda_0$ yields the vortex shedding frequency.
The remaining eigenvalues correspond to higher-order harmonics.
We use the results obtained with the exact DMD on the clean data as a reference.
All variants capture the leading eigenvalue well, regardless of the noise level.
However, there are differences in the harmonics.
This behavior is expected because the noise bias/dampening is more substantial for low energy modes \cite{dawson2016}.
Remarkably, the ADAM/POD variant recovers all eigenvalues almost perfectly, regardless of the noise level.
For clarity, we have omitted the ADAM markers in figure \ref{fig:cylinder_eigvals} because the results were nearly identical to ADAM/POD.
While the bias in the exact DMD increases with the noise level, the VarPro variant does not capture the third and fourth harmonics very well for the investigated SNR.
The results in figure \ref{fig:cylinder_eigvals} are for $r=20$, but $r=15$ and $r=25$ show similar trends and are available in the online material.

\subsection{Transonic shock buffet on a swept wing}
\label{sec:buffet}

Transonic shock buffet on airfoils and wings is a compressible flow phenomenon characterized by the unsteady interaction between shock motion and boundary layer separation.
A shock forms on the airfoil's suction side at positive angles of attack under transonic conditions.
Beyond the critical angle of attack, the boundary layer downstream of the shock begins to separate, and the shock moves toward the trailing edge.
The boundary layer eventually reattaches as the shock moves, and the shock motion is reversed.
When the shock returns to its original position, the process starts again.
For swept wings, this chordwise back-and-forth motion is superimposed with spanwise oscillations, which propagate toward the wing tip.
Together, they form a wavy motion pattern on the wing's suction side, typically called buffet cells.
The periodic interaction between the shock and the boundary layer can lead to a strong periodic interaction between the fluid and the structure, called buffeting.
Buffeting is highly undesirable because it reduces flight comfort and, in the worst case, can compromise the aircraft's structural integrity.
However, flying at high Mach numbers is desirable because of the reduced fuel consumption per flown mile.
Therefore, much research is directed toward understanding unsteady flow phenomena and extending the current flight envelope.

The research unit FOR 2895 investigates the shock buffet phenomenon on the XRF-1 model under realistic flight conditions with respect to the Reynolds number, Mach number, and angle of attack.
Lutz et al. \cite{lutz2022} provide a detailed overview of all measurement techniques and flow conditions investigated.
Moreover, Waldmann et al. \cite{waldmann2023} extensively analyze a significant portion of the available optical surface pressure measurements, pressure and acceleration sensor data, and complementary aerodynamic performance data.
Here, we focus on the DMD analysis of a single unsteady pressure-sensitive paint (PSP) \cite{yorita2023} measurement at Reynolds number $Re_\infty = U_\infty c_m/\nu_\infty=12.9\times 10^6$, Mach number $Ma_\infty=U_\infty/a_\infty = 0.84$, and angle of attack $\alpha = 4^\circ$ ($U_\infty$ - free stream velocity, $c_m$ - mean chord, $\nu_\infty$ - free stream kinematic viscosity, $a_\infty$ - free stream speed of sound). 
This flow condition places the data well within the buffet regime.
The images of the surface pressure coefficient $c_p$ have a spatial resolution of $465\times 159$ pixels.
We analyze a sequence of 1000 snapshots sampled at $2000Hz$.
The 3D buffet is characterized by frequencies at Strouhal numbers $S_r = fc_m/U_\infty$ in the range $0.2 < S_r < 0.6$.
Assuming $S_r \approx 0.4$, the dataset extends over 218 buffet cycles and resolves each cycle with roughly 4.6 snapshots. 

\begin{figure}[htbp]
    \centering
    \includegraphics[width=0.6\textwidth]{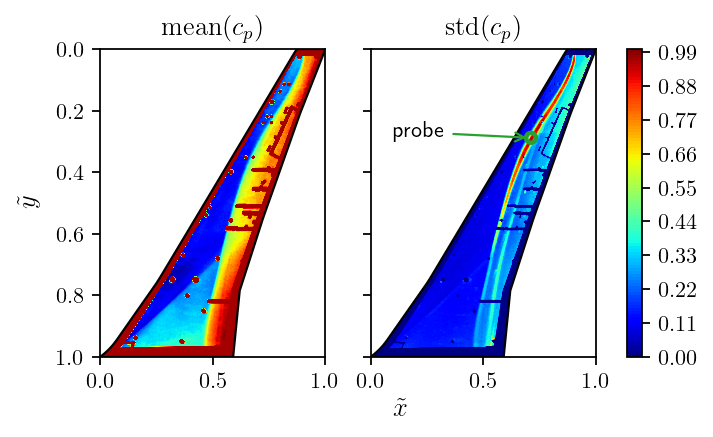}
    \caption{Wing buffet example: temporal mean and standard deviation of $c_p$ scaled to the range $[0, 1]$; $Re_\infty =12.9\times 10^6$, $Ma_\infty = 0.84$, $\alpha = 4^\circ$.}
    \label{fig:cp_mean_std}
\end{figure}

Figure \ref{fig:cp_mean_std} shows the temporal mean and standard deviation of the surface pressure.
The main shock front extends from the wing root to the tip.
This shock oscillates strongly in the region between one third of the span and the tip.
The lambda shock near the root is relatively steady.
Due to the difficult measurement conditions and the embedded surface sensors, some artifacts appear in the recording.
To reduce the impact of these artifacts, we masked out the wing edges and all other sensor locations by setting the corresponding $c_p$ values to zero.
In addition, the standard deviation shows a second elongated region with large pressure fluctuations along the span, starting near the root and extending up to two-thirds of the span (downstream of the mean shock location).
These fluctuations are due to an optical effect introduced by the measurement technique and are irrelevant to the analysis.

The analysis of this dataset is particularly challenging due to low SNR and measurement artifacts that cannot be completely removed despite extensive post-processing efforts.
We employ the ADAM/POD variant with a rank parameter of $r=70$, a batch size of $N_\mathrm{batch}=32$, the default initial learning rate $\alpha_0 = 10^{-3}$, and the default training-validation split of $75/25\%$.
Batch training speeds up the optimization significantly.
Other tested values of $N_\mathrm{batch} = {16, 64, 128}$ had a similar positive effect.
The rank truncation is chosen based on the singular value hard thresholding algorithm \cite{gavish2014}.
The training-validation split is particularly important in this analysis because strong overfitting occurs otherwise (the validation loss increases strongly while the training loss continues to decrease).

To mitigate overfitting and reduce the number of modes, we also add Tikhonov regularization \eqref{eq:f_reg} to the loss function.
To calibrate the hyperparameter $\gamma_0$, we follow the approach suggested by Jovanovi\'c et al. \cite{jovanovic2014} and measure the cardinality of the mode amplitudes $\mathrm{card}(\mathbf{b}_r)$ as well as the performance loss over the range $0 \leq \gamma_0 \leq 0.1$.
To numerically approximate the cardinality, the mode amplitudes $b_i < 10^{-2}$ are set to zero.
Similar to reference \cite{jovanovic2014}, the performance loss is defined as:
\begin{equation}
    \label{eq:perf_loss}
    \Pi_\mathrm{loss}(\gamma_0) = \frac{E_\mathrm{rec}(\gamma_0)}{E_\mathrm{rec}(\gamma_0^\ast)} \times 100\%, 
\end{equation}
where $\gamma_0^\ast$ is the sparsity parameter that gives the lowest reconstruction error.
The reconstruction error is computed on the full dataset (training and validation data).

\begin{figure}[htbp]
    \centering
    \includegraphics[width=0.8\textwidth]{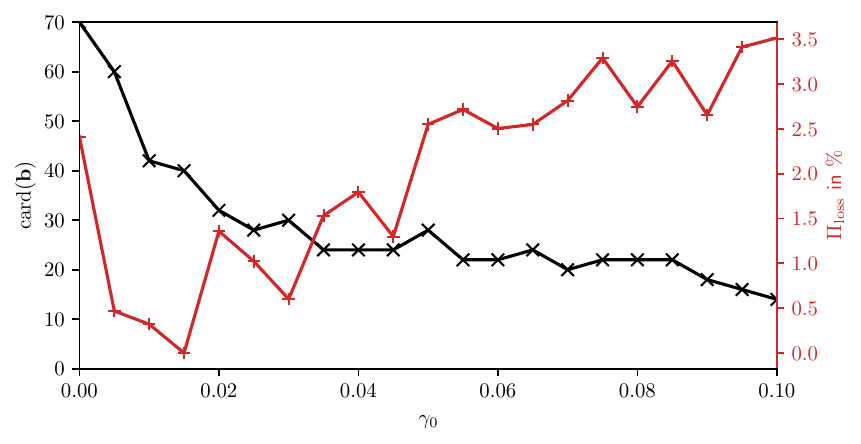}
    \caption{Wing buffet example: effect of the sparsity weight $\gamma_0$ on the cardinality of the DMD amplitudes and the performance loss \eqref{eq:perf_loss}.}
    \label{fig:wing_sparsity}
\end{figure}

Figure \ref{fig:wing_sparsity} shows the influence of the regularization on cardinality and performance loss.
Interestingly, the lowest reconstruction error is not achieved for $\gamma_0 = 0$, but there seems to be a sweet spot around $\gamma_0 \approx 0.01$.
Increasing the regularization further leads to a decrease in reconstruction accuracy.
However, the overall change in the performance loss is only a few percent in the range studied.
The cardinality drops rapidly as soon as the regularization is introduced and begins to level off at $\gamma_0 \geq 0.03$.
Note that only a single run was performed for each value of $\gamma_0$.
Due to the stochastic nature of the optimization, there is a small dependence of the results on the seed value.
However, the trends described above are persistent, so we did not find it necessary to compute convergence statistics.
The subsequent analysis shows results obtained with $\gamma_0=0.03$, balancing sparsity and accuracy.

\begin{figure}[htbp]
    \centering
    \includegraphics[width=0.8\textwidth]{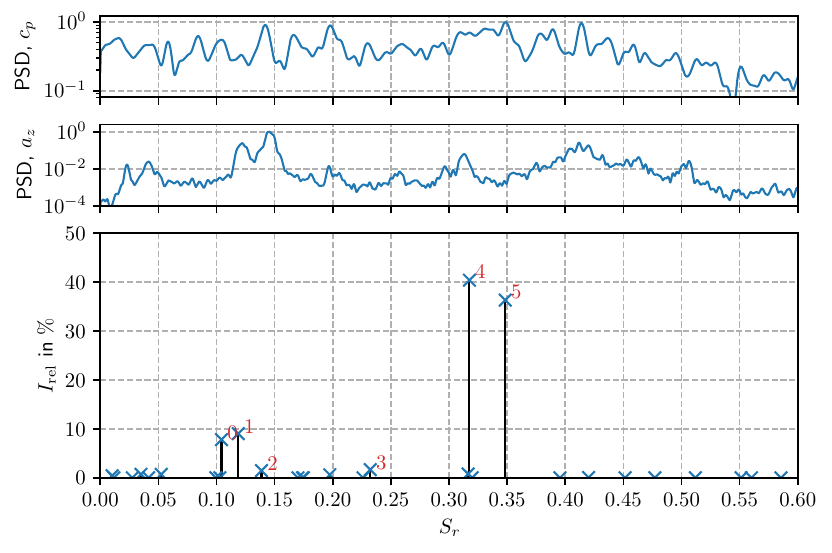}
    \caption{Wing buffet example: comparison of the DMD spectrum (bottom) based on the relative importance $I_\mathrm{rel}$ from equation \eqref{eq:imp_rel}, with the PSD of the vertical wing tip acceleration $a_z$ (center) and the PSD of a $c_p$  time trace at a spatial location highlighted in figure \ref{fig:cp_mean_std} (top). Relevant DMD modes are enumerated. The PSD values are normalized with their respective maximum values.}
    \label{fig:wing_spectrum}
\end{figure}

Figure \ref{fig:wing_spectrum} shows the DMD spectrum complemented by two power spectral density distributions (PSD) for comparison.
We use the integral criterion \eqref{eq:int_criterion} to characterize the importance of the DMD modes/eigenvalues.
To make the importance even more interpretable, we provide relative values computed as:
\begin{equation}
    \label{eq:imp_rel}
    I_{\mathrm{rel},i} = \frac{2I_i}{\sum_{j=0}^{r-1}I_j} \times 100\%
\end{equation}
for mode $i$.
The factor 2 accounts for the fact that each mode has a conjugate complex pair.
The modes can be divided into two groups, one in the range $0.1\leq S_r\leq 0.15$ and one in the range $0.3\leq S_r \leq 0.35$.
The two high-frequency modes, numbered 4 and 5, dominate the pressure fluctuation and together contribute $\approx 79\%$ according to their relative importance.
The second group, numbered 0, 1, 2, contributes $\approx 19\%$.

The PSD of the pressure signal at the probe location marked in figure \ref{fig:cp_mean_std} is relatively flat due to a low SNR.
The maximum PSD appears at $S_r=0.35$, which corresponds perfectly to the frequency of DMD mode 5.
Peaks with similar PSD are visible for Strouhal numbers of $S_r = \lbrace 0.14, 0.2, 0.42 \rbrace$.
Of these peaks, only $S_r=0.14$ has a corresponding DMD mode.
We remark that the $c_p$ signal is taken from the same PSP measurement.
Unfortunately, the signals from the other surface pressure transducers are unusable for this analysis because they are outside the shock oscillation region.

The accelerometer signal $a_z$ quantifies the vibrations of the wing tip in the direction of gravity.
The PSD of the signal has a maximum at $S_r=0.14$ and another smaller peak around $S_r=0.12$.
These frequencies are also matched by DMD modes 1 and 2.
We consider this agreement as an indication of a possible coupling between structural vibrations and the pressure oscillations encoded by the low-frequency modes $0.1\leq S_r\leq 0.15$.

Compared to the other two PSD distributions, the cleaner and more defined DMD spectrum is a clear example of the advantages of global spectral methods for problems with spatio-temporal variations.

\begin{figure}[htbp]
    \centering
    \includegraphics[width=0.8\textwidth]{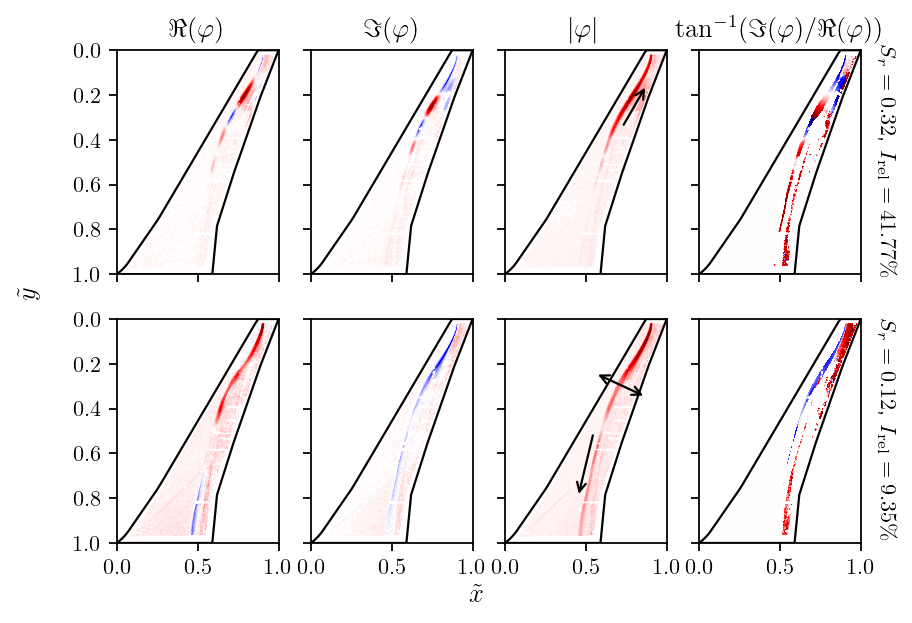}
    \caption{Wing buffet example: modes corresponding to the frequencies marked with 4 and 1 in figure \ref{fig:wing_spectrum}. All fields are scaled to the range $[-1, 1]$. Blue and red colors indicate negative and positive values, respectively.}
    \label{fig:wing_modes}
\end{figure}

The DMD modes presented in figure \ref{fig:wing_modes} provide further insights into the physical phenomena encoded by the two groups of modes.
We visualize only the most dominant mode from each group because the modes within each group are qualitatively similar.
For each mode $\mathbf{\varphi}$, the real part $\Re (\mathbf{\varphi})$, the imaginary part $\Im (\mathbf{\varphi})$, the absolute value $|\mathbf{\varphi}|$, and the phase angle $\mathbf{\phi}$ are shown.
The phase angle is defined as:
\begin{equation}
    \mathbf{\phi}=\tan^{-1}\left( \frac{\Im(\mathbf{\varphi})}{\Re(\mathbf{\varphi})}\right).
\end{equation}
The real and imaginary parts of mode 4 clearly show the buffet cells extending from mid-span to tip.
In the phase angle visualization, we set regions with low $\mathrm{std}(c_p)$ to zero for better visibility.
There are still many artifacts, but the phase angle distribution in the shock region is relatively smooth.
The direction of propagation of the buffet cells can be inferred from the slope of the phase angle in the spanwise direction.
Regions with small phase angles lag behind regions with large phase angles.
Hence, the propagation direction is toward the negative slope (from red to white to blue in figure \ref{fig:wing_modes}), showing that the buffet cells are moving toward the wing tip. 
The propagation speed $U_p$ in some direction $s$ can be approximated as \cite{poplingher2019}:
\begin{equation}
\label{eq:prop_speed}
    U_p = \frac{2\pi f}{\Delta\phi / \Delta s},
\end{equation}
and yields a speed of $U_p\approx 0.22U_\infty$ for the buffet cells (propagation toward the tip).
This value is in the expected range of $0.2U_\infty \leq U_p \leq 0.3U_\infty$ found in the literature \cite{waldmann2023}.
Note that we extract the phase angle slope by a least-squares fit along the shock front, so there is a small sensitivity to the definition of $s$ in \eqref{eq:prop_speed}.
The corresponding wavelength with respect to the mean chord is $l=U_p/f\approx 0.7c_m$.
The propagation speed and wavelength of mode 5 are similar.

Unlike the buffet modes, the low-frequency modes extend over nearly the entire span.
The modes encode strong pressure fluctuations from the midspan to the tip and weaker shock oscillations between the root and the midspan; see $|\mathbf{\varphi}|$ in figure \ref{fig:wing_modes}.
Unfortunately, the phase angle of mode 1 is very noisy.
However, a visual inspection of the reconstructed mode (a reconstruction computed using only a single mode) shows that the shock oscillates along the chord from the midspan to the tip, while the shock oscillations in the wing's inner part propagate toward the root.
Evaluating the propagation speed of these waves yields $U_p=0.24U_\infty$ (toward the root), which is close to the speed of the buffet cells.
Interestingly, the resulting wavelength is almost exactly twice the mean chord, i.e., $l=2c_m$.
We suspect that the low-frequency modes may be related to wing bending and/or twisting modes.
A structural analysis of the wing under wind-on conditions would be beneficial to gain further insight.
The inboard traveling shock waves have already been observed before by Massini et al. \cite{masini2020} on the Common Research Model.
The authors also suspected a coupling to structural vibrations but did not provide concrete evidence.
Ultimately, we cannot confirm this hypothesis, but the agreement between DMD and accelerometer PSD, as well as the wavelength of $2c_m$, provide strong evidence for a structural coupling.

Finally, we would like to emphasize that the results presented in this section are only possible due to the robustness of the proposed DMD methodology.
We have repeated the analysis with rank truncation values of $r=\lbrace 90, 120\rbrace$ and found nearly identical results.
We did not obtain similar results with the exact DMD or VarPro variants.
The exact DMD yields strongly damped modes with the largest amplitudes in the range $S_r \leq 0.15$.
The variant also fails to identify the buffet, which is an essential flow feature that can even be observed by visual inspection of the $c_p$ snapshots.
The VarPro variant places the buffet in the correct frequency range but identifies strongly growing dynamics, likely due to overfitting.
A comparison of the eigenvalues of the three DMD variants is provided in the appendix.

\section{Conclusion}
\label{sec:conclusion}
In this paper, we introduce a new methodology to compute the optimized DMD.
Our method employs algorithms from the DL community to solve nonlinear optimization problems to obtain the eigendecomposition of the DMD operator.
The new approach comes at an increased computational cost, but significantly improves accuracy and robustness compared to existing DMD variants.
The robustness and accuracy are persistent for all example problems and hyperparameter variations investigated here.
We believe that this reliability will be of great help to practitioners working with challenging datasets such as the PSP measurements presented in section \ref{sec:buffet}.
Thanks to the robustness and accuracy of the proposed DMD approach we are able to accurately analyze the complex wing buffet measurements and identify the relevant modes.
The approach is extremely flexible and requires very little effort to explore other definitions of the DMD operator.
Extensions to parametric, control-oriented, or probabilistic DMD variants are also possible.
Such variants would also make DMD more attractive as a choice for predictive reduced-order modeling and could provide an alternative to less interpretable approaches based on neural networks \cite{zahn2023}.


\section*{Acknowledgment}
The authors gratefully acknowledge the Deutsche Forschungsgemeinschaft DFG (German Research Foundation) for funding this work in the framework of the research unit FOR 2895 under the grant WE 6948/1-1. The authors would like to thank the Helmholtz Gemeinschaft HGF (Helmholtz Association), Deutsches Zentrum für Luft- und Raumfahrt DLR (German Aerospace Center) and Airbus for providing the wind tunnel model and financing the wind tunnel measurements as well as public support to mature the test methods applied by DLR and ETW.

\appendix
\section*{Appendix}
\begin{figure}[htbp]
    \centering
    \includegraphics[width=0.8\textwidth]{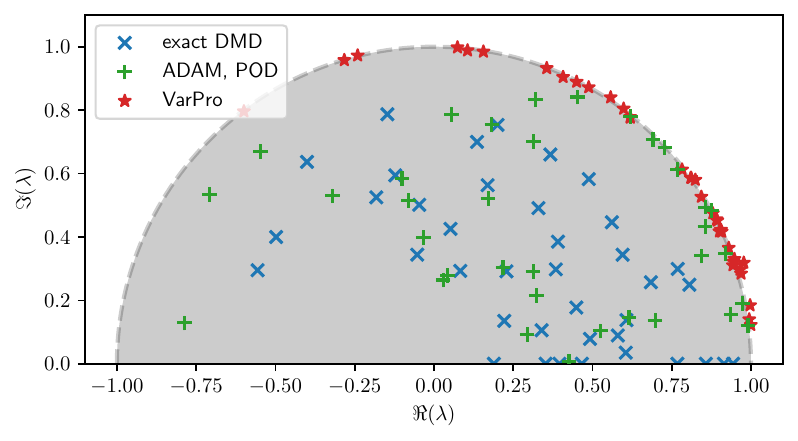}
    \caption{Wing buffet example: comparison of eigenvalues $\mathbf{\lambda}$ obtained with exact DMD, VarPro, and ADAM/POD.}
    \label{fig:wing_eigval_comp}
\end{figure}

Figure \ref{fig:wing_eigval_comp} shows a comparison of $\mathbf{\lambda}_r$ obtained for the PSP dataset with exact DMD, VarPro, and ADAM/POD.
All variants are optimized based on the first 750 snapshots.
The exact DMD identifies only strongly decaying dynamics, which is a consequence of the PSP data's low SNR.
The VarPro algorithm identifies a significant number of growing dynamics, especially near $\Im(\lambda)\approx 0.3$.
The ADAM/POD variant identifies a small number of modes that are persistent over time.
These are the modes described in section \ref{sec:buffet}.
Note that the ADAM/POD results are obtained without regularization, i.e., $\gamma_0=0$, to provide a fair comparison.

\printbibliography[title=References, heading=bibliography]

\end{document}